\documentstyle[12pt,epsf]{article}

\begin{document}

\begin{center}

{\Large\bf Searching for  self-similarity in switching time and turbulent cascades in ion transport
through a biochannel.  A time delay asymmetry}

\vspace*{1cm}

{\large Marcel Ausloos\footnote{Email address: marcel.ausloos@ulg.ac.be}} \\
\vspace*{2mm} SUPRATECS \& GRASP, Institute of Physics B5,\\ University of
Li\`ege, B-4000 Li\`ege, Euroland

\vspace*{5mm}

{\large Kristinka Ivanova\footnote{Email address: kristy@essc.psu.edu}}

\vspace*{2mm}

Pennsylvania State University, University Park, PA 16802, USA

\vspace*{5mm}

{\large Zuzanna Siwy\footnote{Email address: siwy@zeus.polsl.gliwice.pl}}

\vspace*{2mm} Materials Research Department, \\ Gesellschaft f\"{u}r
Schwerionenforschung, 64291 Darmstadt, Germany

\vspace*{2mm}

and

\vspace*{2mm}

Department of Physical Chemistry and Technology of Polymers,\\ Silesian
University of Technology, 44-100 Gliwice, Poland \\present address: Department of
Physics, \\Univ. of Florida, Gainesville, FL 32611, USA

\end{center}

\vspace*{1cm}

\noindent

{\bf Abstract:}

The process of ion transport through a locust potassium channel is described by
means of the Fokker-Planck equation (FPE). The deterministic and stochastic
components of the process of switching between various conducting states of the
channel are expressed by two coefficients, $D^{(1)}$ and $D^{(2)}$, a drift and a
diffusion coefficient, respectively.  The FPE leads to a Langevin equation. This
analysis reveals beside the well known deterministic aspects a turbulent, cascade
type of  action. The (noisy-like) switching between different conducting states
prevents the channel from staying in one, closed or open state. The similarity
between the hydrodynamic flow in the turbulent regime and hierarchical switching
between conducting states of this biochannel is discussed. A non-trivial
character of $D^{(1)}$ and $D^{(2)}$ coefficients is shown, which points to
different processes governing the channel's action, asymetrically depending on
the history of the previously conducting states. Moreover, the Fokker-Planck and
Langevin equations provide information on whether and how the statistics of the
channel action change over various time scales.

\noindent



\vspace*{1cm}

\section{Introduction}

Ionic transport through biological membranes is one of the basic phenomenological
processes necessary for maintaining a system alive (Hille, 1992; Schrempf et al.,
1995). The mechanism, even after so many years and so many intense investigations
is not yet fully understood nor even well known. A full picture should consist of
combinations of causes with mechanical, electrical, thermodynamic and chemical
origins (Hille, 1992; Cha et al., 1999). Ion channels are thought to be made of
flexible transmembrane proteins (Hille, 1992).   The ion transport through a
biological channel is certainly a process of very mixed and complex nature, being
governed by the noise as well as deterministic forces (Colquhoun and Hawkes,1995)
coming e.g. from external or endogenous field(s). A serious step forward was
done by Bezrukov and Winterhalter who put forward the conjecture that it is the
motion of channel structural constituents that induces channel switching between
the open and closed states (Bezrukov and Winterhalter, 2000), as noise does in
stochastic resonance processes (Benzi et al., 1981). An evidence for this
hypothesis has been experimentally provided recently on the basis of artificially
controlled nanopores (Siwy and Fuli\'{n}ski,
2002)\footnote{http://focus.aps.org/story/v10/st19}. Indeed it can be now clear
that a nanopore which possesses mobile elements exhibits scale dependent ion
current fluctuations while the nanopore with smooth and stable walls produces
stable ion current signals. The (stochastic) movement of the channel structural
constituents seems therefore to be a dominant factor determining the channel
behaviour. Moreover, those studies showed that the power spectrum of the ion
current fluctuations is inversely proportional to a power of the frequency, which
points to, but does not prove, the self-similarity in time of processes
underlying the channel switching. Multiple exponentials in the lifetime
distributions themselves indicate high degrees of correlations (Colquhoun and
Sigworth, 1995).

Our knowledge on the random and/or deterministic character of those processes is
however very limited. More or less sophisticated techniques are in fact used to
identify features embedded in single-channel recordings (Fredkin and Rice, 1992;
Mercik, et al., 1999; Wagner and Timmer, 2000; Mercik and Weron, 2001; Siwy et
al., 2001; Siwy et al., 2002). A difficulty stems in the fact that the whole
membrane structure is of nanometer scale and can hardly  be locally probed and
the tiny electrical current measured through a single channel can reach
$10^{-11}$ ampere order of magnitude. Yet, in the patch clamp technique (Neher
and Sakmann, 1976) when constant voltages are applied, the ion current switches
between regimes, be they composed of discrete or not levels is still an open
question, - the switching often resembling some dichotomous noise. Let it be
emphasized that even though the observed ion current fluctuations look random, a
non-Markovianity, i.e. causality, of the process seems proven in a number of
cases.

One of the especially extensively examined systems is a voltage-gated channel in
a locust muscle (BK channel) (Gorczy\'{n}ska et al., 1996). It reveals well
distinguished conducting regimes: high values of ion current corresponding to the
open state(s), while low currents point to channel closed state(s). In each
regime, the measured currents surely fluctuate around a certain conductance
level, but the fluctuation distributions, being discrete or not, of finite range
or not,  are far from universal but are of not the main concern here. We focus
our attention on the {\it features, which govern the switching} between those
(two ensembles of) states, and study the time series of their durations. The
signal of ion current through a single BK channel has been previously subjected
to the detailed analysis by the autocorrelation function (Mercik et al., 1999),
the Hurst exponent for the long range correlations (Mercik, et al., 1999; Mercik
and Weron, 2001; Siwy et al., 2001) as well as checking Markovianity on the basis
of the Smoluchowski-Chapman-Kolmogorov equation (Fuli\'{n}ski et al., 1998). The
existence of correlations and deterministic components have been found by all
those methods. However it was also shown that a state memory effect is present
not only between successive conducting states of the channel but also
independently within the closed and open states themselves (Siwy et al., 2001).
Thus the next step should be to characterize the random aspects in closing and
opening of such ion channels in some way.

One rigorous way to sort out the noise from the deterministic components is to
examine in details correlations at $different$ time lag scales through the so
called master equation, i.e. the Fokker-Planck equation (and derivation of the
subsequent Langevin equation) for the probability distribution function ($pdf$)
of time and space signal increments (Gardiner, 1983; Friedrich et al., 2000).
This theoretical approach, based on $rigorous$ statistical principles, is often
the first step in sorting out the $best$ model(s). In this paper we would like to
derive such an FPE, directly from the experimental data of ion current
fluctuations in the BK channel, whence describing the evolution of e.g. dwell
time differences as a function of the  number of interval events (number of
closed events for differences in closed dwell times, number of open events for
differences in open dwell times, etc.). In order words we examine series of
dwell-times durations consisting of successive open ($o$) and closed ($c$) states
$T_{o,c}(i_{o,c})$, as well as the series of durations of closed $T_{c}(i_{c})$
and open states $T_{o}(i_{o})$, treated separately.

Following an approach similar to (Friedrich et al., 2000; Renner et al., 2001;
Ausloos and Ivanova, 2003; Ivanova et al., 2004) we show that the FPE can be well
derived in terms of the first (two) moments of the experimental data probability
distribution function $p(\Delta T, \Delta n)$, where $\Delta T$ is an {\it
increment  of the closed or dwell-time} series, and $\Delta n$ stands for the
{\it time delay} between the specific states. We would like to emphasize that the
method is model independent. Moreover the technique allows examination of long
and short time scales {\it on the same footing}. The discussion of the results
seem to induce some thinking about new features of correlations in ionic
transport. Such an approach is in fact in line with unclassical ion transport
models, e.g. based on fractal ideas for the geometry of such membranes (Roncaglia
et al., 1994; Bassingthwaighte et al., 1994; Liebovitch and Todorov, 1996), 
though multifractal considerations might be ... timely. From the start let us
recall however that a multifractal process in time can exist on a non-fractal
geometrical network.

The solution of the FPE yields the probability distributions with high accuracy.
Furthermore, the so found analytical form of both drift $D^{(1)}$ and diffusion
$D^{(2)}$ coefficients appearing in the FPE  has a simple physical
interpretation, reflecting the influence of the deterministic and random forces
on the examined ion transport process. For instance deterministic forces can be
thought to be derived from potential gradients in the case of physical or
chemical fields, and  originate from thermal (or  other physical) noise in the
case of stochastic forces.

Additionally, the time lag fluctuation probability distributions and the
diffusion coefficient enable to trace the existence of hierarchy or cascade
processes, underlying the switching phenomena. The analogy between the FPE for
the ion transport through membranes can thus be briefly compared with the similar
one performed for a hydrodynamic flow at the turbulent regime onset. It evokes
nontrivial mechanisms to be included in realistic models.

\section{Experimental data and variables to be analyzed}

We hereby discuss a long data set of ion current recorded through a single
large-conductance locust potassium channel (BK channel), as described in
(Gorczy\'{n}ska, 1996; Fuli\'{n}ski et al., 1998). The so published and selected
for its length ion current data consists of 250~000 points separated by 0.1 ms,
thus the total duration of the time series is 25 s. The probability density
function of the ion current has been found to be distinctly bimodal, with a well
marked threshold at $I^*=5.6\pm0.2$ pA, as described in (Mercik et al., 1999).
The value of $I^*$ separates therefore $two$ modes of the ion current action. A
given ion current point is categorized as belonging to an open state if $I>I^*$,
otherwise it belongs to the closed state time series. This procedure led to
formation of the dwell-time series of all  durations of open and closed states,
starting with open state duration, $T_{o,c}(i_{o,c})$, ($i_{o,c}$=1, ..., 30604)
as well as two series of durations of closed $T_{c}(i_{c})$, ($i_{c}$=1...15302)
and open states $T_{o}(i_{o})$, ($i_{o}$=1...15302), treated separately. We have
focused on the analysis of the dwell-times series.

Examination of the fluctuations of a time series can be performed via studies of
the intervals' fluctuations. The intervals can be expressed by calculating
returns, log-returns, or increments  (Ausloos and Ivanova, 2003). Here, we have
used the increments

\begin{equation}  \Delta T_{o,c}(\Delta n_{o,c})=T_{o,c}(j_{o,c}+\Delta n_{o,c})-
T_{o,c}(j_{o,c}) , \qquad  j_{o,c}=1,...,(i_{o,c}-1), \end{equation}

\vspace*{5mm}

\begin{equation} \Delta T_{c}(\Delta n_c)=T_{c}(j_{c}+\Delta n_{c})- T_{c}(j_{c})
, \qquad j_{c}=1,...,(i_{c}-1), \vspace*{5mm} \end{equation}

\begin{equation} \Delta T_{o}(\Delta n_o)=T_{o}(j_{o}+\Delta n_{o})- T_{o}(j_{o})
, \qquad j_{o}=1,...,(i_{o}-1), \end{equation}

where $\Delta T$ is a time  increment  of the time state duration, and $\Delta n$
stands for the time delay between the specific states. Derivation of the
Fokker-Planck equation presented in the next Section has been based on single and
joint probability density functions (pdf) of $\Delta T_{o,c}(\Delta n_{o,c})$,
$\Delta T_{c}(\Delta n_c)$ and $\Delta T_{o}(\Delta n_o)$.

The analysis has been performed for the experimental data and their surrogates
created by random shuffling of the data positions (Bassingthwaighte et al.,
1994). This enabled us checking the significance of obtained results.

\section{Results and discussion}

\subsection{Probability density functions}

Figures 1-3 show the probability density functions
(pdf) of the dwell-time increments for the $open-closed$, $closed$ and $open$
series. We have used different discrete bins to obtain interval contents of
similar length, needed for the statistical tests. The corresponding shuffled
($sh$) cases to be used for statistical tests are not shown for the sake of
space. Note that the pdfs of the $closed-state$ durations and of the time series
of $open$ and of $closed$ dwell-times are qualitatively similar (for open states
only the $\Delta T$ range is significantly shorter). Moreover, all pdf's show a
surprisingly $high$ probability for $large $changes of the states duration (up to
300 ms !), so called extreme events. Additionally, in the examined time ranges,
the obtained pdfs change their character very weakly, which is in a very good
agreement with information given by the detrended fluctuation analysis (DFA)
studies. We would like to note that the existence of positive correlations in all
series of dwell-times has been previously shown in the same analyzed range of the
lag $\Delta n$ (Siwy et al., 2001; Siwy et al., 2002) as the one used here.

More detailed information about the correlations present in the time series is
given by joint pdf's, depending on $N$ variables, i.e. $p^N (\Delta T_1,\Delta
n_1;...;\Delta T_N,\Delta n_N)$. We started to address this issue by determining
the properties of the pdf for $N=2$, i.e. $p(\Delta T_{2},\Delta n_{2}; \Delta
T_{1}, \Delta n_{1})$. Contour plots for the equal values of $\Delta T$'s offer a
clear way of presenting the results. The tilted and asymmetric (triangular)
character of the pdf's (Fig. 4(a-c)) for the examined series is quite
unusual and points to the statistical dependence (correlation) between the
increments in all examined time series. $Negative$ increments especially are
characterized by very $high$ correlation, which confirms a persistency of the
time series towards (negative) changes: {\it the decrease of dwell-times
durations is most probably followed by another decrease in the next time step}.
For the $positive$ increments the correlation is more complex and shows $two$
types of correlations, corresponding to an asymmetry in the peaked distributions.
This effect is especially distinct for the series of $\Delta T_{o,c}$ and $\Delta
T_{c}$. Note that these correspond to the largest $\Delta T$ ranges, thus to
extreme events.

Other methods such as the hidden Markov model analysis  incorporate filtering and
detection of sub-threshold events on a background of noise (Colquhoun and
Sigworth, 1995; Wagner and Timmer, 2000). This maybe prone to missing extreme and
unusual events. Our approach, in contrast, considers that we can simply
distinguish two well defined types of states. This might be a simplification, and
we are aware that the number of states might be continuous rather than discrete.
One might discretised (but under which criteria?) the pdf's and look for
correlations between sub-pdf's. It is not obvious whether such an arbitrariness
is presently useful. We have examined elsewhere (in a quite different context)
whether such a distribution could be characterized (Ausloos and Ivanova, 2003).
Application of the technique to the present case is left for further work. The
quantity of data in the present case should make the result fascinating.

\subsection{Kramers-Moyal coefficients and Fokker-Planck equation}

As the next step the general case of $p^N$($\Delta T_1$,$\Delta n_1$;...;$\Delta
T_N$,$\Delta n_N$) has been considered. If the increments follow a Markov process
a serious simplification occurs\footnote{The $N$-point pdf $p^N$ is generated by
the product of conditional probabilities, e.g. $p(\Delta T_{i+1},\Delta n_{i+1} |
\Delta T_{i},\Delta n_{i}) = p(\Delta T_{i+1},\Delta n_{i+1} ; \Delta
T_{i},\Delta n_{i}) / p(\Delta T_{i},\Delta n_{i})) $ for  $ i = 1, ... N-1$}
rigorously described by the Chapman-Kolmogorov equation

\begin{equation} p(\Delta T_{2},\Delta n_{2} | \Delta T_{1},\Delta n_{1}) = \int
d(\Delta T_{i}) p(\Delta T_{2},\Delta n_{2} | \Delta T_{i},\Delta n_{i}) p(\Delta
T_{i},\Delta n_{i} | \Delta T_{1},\Delta n_{1}) \end{equation}

\noindent

Considering the limiting case of $\Delta n$ tending to zero, the (continuous time
"$\tau$") differential form of the Chapman-Kolmogorov equation yields a master
equation, giving the Fokker-Planck equation (Gardiner, 1983; Risken, 1984). The
moments $M^{(k)}$ of the probability density distribution, the so called
Kramers-Moyal coefficients, enable estimation of the drift $D^{(1)}$ and
diffusion $D^{(2)}$ coefficients,\footnote{$D^{(k)} \simeq  M^{(k)}/k!$ }
respectively, such that the FPE reads

\begin{equation} \frac {d}{d\tau} p (\Delta T, \tau)= [ -
\frac{\partial}{\partial \Delta T } D^{(1)} ( \Delta T , \tau ) +
\frac{\partial}{\partial ^2 \Delta T^2} D^{(2)}(\Delta T,\tau)] p (\Delta T,
\tau) \end{equation}

For completeness, in Fig. 5, we show the equal probability contours for
the open-closed state conditional probability. The experimental (so called
''empirical'') pdf is compared with the pdf resulted from the Chapman-Kolmogorov
equation for different triplets of $\Delta n_2<\Delta n_i<\Delta n_1$. As shown
in Fig.~5 we have obtained a fairly good agreement between the two pdf.
Note however that the joint and conditional pdf differ in their approach to the
assessment of the disappearance of a correlation axis for small $\Delta T$ in the
latter.

Figures~6-8 show the dependence of $D^{(1)}$ and
$D^{(2)}$ on the increments of dwell-times $\Delta T$.  For the open-closed and
open states the dependence of the coefficients is $asymmetric$ with respect to
$\Delta T$. For positive $\Delta T$, it is linear for $D^{(1)}$ and quadratic for
$D^{(2)}$. Practically no dependence on $\Delta T$ was found for $\Delta T<0$.
For open states the dependence of the diffusion coefficient on $\Delta T$ is
quadratic; the drift coefficient is asymmetric with respect to $\Delta T$ and
shows a linear dependance.

\vspace*{5mm}

\[ D^{(1)}_{o,c}\propto\left\{ \begin{array}{ccl} -1.1139 \Delta T + 0.1684& {\rm
for } & \Delta T>0 {\rm ms}\\ 0.0081 \Delta T - 0.3241 & {\rm for } & \Delta T<0
{\rm ms}\\ \end{array} \right. \]

\vspace*{5mm}

\[ D^{(1)}_c\propto\left\{ \begin{array}{ccl} -0.8806 \Delta T + 0.1847& {\rm for
} & \Delta T>0 \, {\rm ms}\\- 0.1740 \Delta T +  0.1284 & {\rm for } & \Delta T<0
\, {\rm ms}\\ \end{array} \right. \]

\vspace*{5mm}

\[ D^{(1)}_o\propto\left\{ \begin{array}{ccl} - 0.9679 \Delta T + 0.3091& {\rm
for } & \Delta T>0 \, {\rm ms}\\ - 0.0701 \Delta T + 0.3139 & {\rm for } & \Delta
T<0 \, {\rm ms}\\ \end{array} \right. \]

\vspace*{5mm}

\[ D^{(2)}_{o,c}\propto\left\{ \begin{array}{ccl} 0.3263 \Delta T^{2} - 3.2864
\Delta T - 4.2590 & {\rm for } & \Delta T>0 {\rm ms}\\ 0.0525 \Delta T + 2.2371&
{\rm for } & \Delta T<0 {\rm ms}\\ \end{array} \right. \]

\vspace*{5mm}

\[ D^{(2)}_c\propto  0.2715 \Delta T^{2} +1.1583 \Delta T +  2.0903  \]

\vspace*{5mm}

\[ D^{(2)}_o\propto\left\{ \begin{array}{ccl}  0.3872 \Delta T^{2} - 0.0948
\Delta T +  0.3531& {\rm for } & \Delta T>0 \, {\rm ms}\\ - 0.0547 \Delta T +
0.4214 & {\rm for } & \Delta T<0 \, {\rm ms}\\ \end{array} \right. \]

Interestingly, such a quadratic scaling of $D^{(2)}$ has also been identified in
the analysis of a fully developed turbulence where the velocity increments were
studied over space scales (Ghashghaie, et al., 1996). Moreover, the quadratic
dependence of $D^{(2)}$ has been found essential for the logarithmic scaling of
intermittency (Ghashghaie, et al., 1996). This statistical behaviour has been
explained in the light of cascade processes from large to small scales. In fluid
flow it is the energy which is transported to the system on large scales and
becomes distributed over smaller scales due to the instability of vortices at a
given scale, thus making the dissipation of energy possible. Ion channels are
perfect systems for the creation of such cascades, i.e. hierarchical processes,
because of the large number and varied types of degree of freedom of the
nonequilibrium process. Protein building the channel performs movements on a very
wide time scale (Liebovitch and Todorov, 1996), which can be seen as
self-similarity in time, and correlation between ion current values and
dwell-states. The hierarchical energy cascade analogy thus fully holds here.
Whether the multifractal spectrum can be deduced is still  an open question
requiring further work.

Finally, we have checked whether the diffusion coefficient for all dwell-times
$D^{(2)}_{o,c}$ is related with diffusion coefficients of open and closed times,
respectively: the following relation is found
$D^{(2)}_{o,c}=(D^{(2)}_c+D^{(2)}_o)/2$. Moreover for all the series examined we
have found that the drift coefficient $D^{(1)}$ is dominant over diffusion
$D^{(2)}$, through the ratio $D^{(2)}/D^{(1)}\Delta T$, $D^{(3)}/D^{(1)}(\Delta
T)^2$, and $D^{(4)}/D^{(1)}(\Delta T)^3$, where $D^{(3)}$ and $D^{(4)}$ are the
third and fourth coefficients of the Kramers-Moyal expansion, corresponding to
$M^{(3)}$ and $M^{(4)}$ respectively.

Additionally, knowing that FPE is equivalent (in Ito sense) to a  Langevin
equation (Risken, 1984), which in the examined case would describe evolution of
each type of $\Delta T$, the following equation has been found

\begin{equation} \frac {d}{d\tau} \Delta T(\tau) = D^{(1)}(\Delta T(\tau),\tau) +
\eta(\tau) \sqrt {{D^{(2)} (\Delta T(\tau),\tau)}}, \end{equation} where
$\eta(\tau)$ is a fluctuating-correlated force with Gaussian statistics; here $<$
$\eta(\tau)$ $\eta(\tau')$$>$ = 2$ \delta (\tau -\tau')$.

\section{Discussion}

The main objective of the paper was to examine the character of ion transport
through a potassium channel by means of a master equation applied to the ion
current time series. We have focused on the processes underlying switching
between different conducting ($o,c$) states of the channel. We aimed at
separation of the deterministic and random components of the process. The derived
Fokker-Planck and Langevin equations of the studied dwell-time series gave a
straight forward answer to the posed question. The drift coefficient contains
information about the $deterministic$ forces influencing the process, while the
diffusion coefficient expresses the noise present in the system (see the Langevin
equation above). For all examined series a clear dominance of the drift component
has been observed. This should have medical and pharmaceutical relevance and
implications.

Scaling of $D^{(1)}$, $D^{(2)}$, and of probability density functions confirmed
rigorously the previous results that the character of $closed$ states
significantly influences the channel behaviour. The $open$ states time intervals
are usually very short, and their pdf is quite narrow whatever the time lag scale
considered. Open pore configurations have been recently well described (Jiang et
al., 2002). The analysis performed here suggests to perform an extra thinking on
and interpretation of the results based on  the time dependence. Notice that
treating ion transport as a superposition of merely two processes, opening and
closing of the channel, the closed state is the slowest part of the process. Its
characteristics govern therefore the overall dynamics of switching between the
states; it seems trivial to state that the ion current is a measure of the
velocity of the ion transport through the channel, but it is worth reemphasizing
because again this might have interesting physiological  causes and practical
impact.

The identified turbulent, cascade type of switching of open states and for the
positive increments of closed dwell times is, as we believe, of crucial
importance for the channel proper function. The intermittency character of this
process (in the sense of Kolmogorov log-normal turbulence model (Frisch, 1995))
promotes the switching between the states, preventing therefore the channel from
staying in one, closed (inactivation) or open (hyperactivation) state (Hille,
1992). We hint that this could be interpreted through the role of noise in forced
(two or multi) level systems (de Oliveira et al., 2003).

We claim that we see an analogy between the turbulent variables, in fluids and
the ion transport case. Hydrodynamic turbulence considered in $spatial$
coordinates corresponds to the ion transport through a channel whose $time$
development has been studied. In turbulence, the laminar periods are interrupted
by turbulent bursts. In ion transport the intermittency is expressed in the
switching between closed and open states. This is tantamount to the periods when
large number of ions is transported through the channel, interrupted by the
periods where the transport is seriously diminished. This leads to the {\it
charge cascade} occurring in time hierarchy, similar to the {\it energy cascade}
in space hierarchy observed in turbulence.

The non-Markovian nature of the closed state dwell time distribution is usually
explained by postulating multiple closed states (the number and distribution of
which are still in question). The interpretation of non-Markovian dwell time
distributions has a bearing on what philosophy of ion channel kinetics is
envisaged, i.e. fractal vs. discrete state Markov models vs. hierarchical energy
landscape,... In the above, we concur with the multifractal view point about the
existence of hierarchical processes as first suggested by the spectral analysis
of the ion current time series (Siwy and Fuli\'{n}ski, 2002). The presence of the
flicker noise pointed to the self-similarity in time of the processes governing
the ion transport. Here, we confirm  this conjecture by showing an unexpected
high probability of large changes of dwell-times durations, revealed by pdf
(compared to a Gaussian pdf).  Recall that the existence of finite non-Gaussian
tails for large events is usually thought to be a result of drastic evolutions
e.g. financial crashes, earthquakes, heart attacks in functioning of an organism,
turbulence and intermittency in fluid flow, .... In ion transport through a
$voltage-sensitive$ channel it may be the mere consequence of exciting the
$voltage-sensor$ i.e. part of the channel, responsible for the response of the
channel to the applied electric field but fundamentally the result from the
complex electrochemical and mechanical fluctuation couplings. We suggest that to
separate the  coupling energies, forces, and to quantify the parameters can be
done by systematically analyzing and considering the various possible cascades,
and their time scales.

\section{Conclusion}

To conclude, we would like to stress that the presented Fokker-Planck and
Langevin equations have been derived through a model$-$free identification of
charge transport trough membranes. They provide the knowledge how the statistics
of the channel action changes over various time scales. For ion channels
$structural$ changes (Jiang et al., 2002) are crucial to every aspect of their
function: ion conduction, gating and pharmacology. The present report shows a
rigorous way to describe the correlation and $time$-hierarchy in ion transport
through channel within a turbulent cascade analogy.

{\it In fine}, the Fokker-Planck equation is often used in the literature to
relate the Brownian motion of selective macroscopic variables (through a Langevin
equation) across some free energy landscape to observables like the gating or
ionic current. The FPE solution reduces to the Boltzmann distribution at
equilibrium, and the noise term satisfies a simple fluctuation-dissipation
theorem. Thus a usual FPE approach to ion channel kinetics might be expected to
lead to simple Gaussian distribution functions for observables. It has been shown
that this is the case for small variations, but the tails of the distributions
(see Figs. 1-3) are not gaussian, and ''extreme'' events are not negligible. Thus
physical models should be improved to describe the tails. This will not be easy.

\newpage


{\small

Ausloos, M. and K. Ivanova. 2003. Dynamical model and nonextensive statistical
mechanics of a market index on large time windows. {\it Phys. Rev. E} 68: 046122
(13 pages).

\vspace*{4mm}

Bassingthwaighte, J.B., L.S. Liebovitch, and B.J. West. 1994. Fractal Physiology.
University Press, Oxford.

\vspace*{4mm}

Benzi,~R., A. Sutera, and A. Vulpiani. 1981. The mechanism of Stochastic
Resonance" {\it  J. Phys.A: Math. Gen.}. 14: L453-L457.

\vspace*{4mm}

Bezrukov,~S.M., and M.~Winterhalter. 2002. Examining Noise Sources at the
Single-Molecule Level: 1/f Noise of an Open Maltoporin Channel, {\it Phys. Rev.
Lett}. 85: 202-205.

\vspace*{4mm}

Cha,~A., G.E.~Snyder, P.R.~Selvin, and F.~Bezanilla. 1999. Atomic scale movement
of the voltage-sensing region in a potassium channel measured via spectroscopy.
{\it Nature}. 402: 809-813.

\vspace*{4mm}

Colquhoun,~D., and A.G.~Hawkes. 1995. The principles of the stochastic
interpretation of ion-channel mechanisms. In: Sakmann B. and E. Neher,  Eds.
Single-Channel Recording. 2 ed. New York: Plenum Press. p 397-482.

\vspace*{4mm}

Colquhoun,~D., and  F.~Sigworth. 1995. Fitting and statistical analysis of single
channel records. In: Sakmann B. and E. Neher,  Eds. Single- Channel Recording.
New York: Plenum. p 483-587.

\vspace*{4mm}

de Oliveira,~ C.R., C. M.~ Arizmendi and J. ~Sanchez. 2003. Multifractal analysis
and randomness of almost periodic forced two-level systems.  {\it Physica A}.
330: 400-408.

\vspace*{4mm}

Fredkin,~ D.R., J.A.~Rice. 1992. Maximum likelihood estimation and identification
directly from single-channel recordings. {\it  Proc.  R. Soc. London B}
249:125-132.

\vspace*{5mm}

Friedrich,~R., J.~Peinke and Ch.~Renner. 2002. How to Quantify Deterministic and
Random Influences on the Statistics of the Foreign Exchange Market. {\it Phys.
Rev. Lett}. 84: 5224-5227.

\vspace*{4mm}

Frisch, U. 1995.  Turbulence: the Legacy of A. N. Kolmogorov. Cambridge:
Cambridge Univ. Press.

\vspace*{4mm}

Fuli\'nski,~A., Z.~Grzywna, I.~Mellor, Z.~Siwy, and P.~N.~R.~Usherwood. 1998.
Non-Markovian character of ionic current fluctuations in membrane channels. {\it
Phys. Rev. E}. 58: 919-924.

\vspace*{4mm}

Gardiner,~C.W. 1983. Handbook of Stochastic Methods. Springer-Verlag, Berlin.

\vspace*{4mm}

Ghashghaie,~S., W.~Breymann, J.~Peinke, P.~Talkner, and Y.~Dodge. 1996. Foreign
Exchange Market - A Turbulent Process? {\it Nature}. 381: 767-770.

\vspace*{4mm}

Gorczy\'nska,~E., P.~L.~Huddie, B.~A.~Miller, I.~R.~Mellor, R.~L.~Ramsey, and
P.~N.~R.~Usherwood. 1996. Potassium channels of adult locust (Schistocerca
gregaria) muscle, {\it Pfl\"uges Arch. Ges. Physiol. Menschen Tiere}. 432:
597-606.

\vspace*{4mm}

Hille,~B. 1992. Ionic Channels of Excitable Membranes.  Sunderland: Sinauer Inc.

\vspace*{4mm}

Ivanova,~K., M.~Ausloos, and H. Takayasu. 2004. Deterministic and stochastic
influences on Japan and US stock and foreign exchange markets. A Fokker-Planck
approach, in :  H. Takayasu, Ed. The applications of econophysics,  Proceedings
of the Second Nikkei Econophysics Symposium.  Berlin: Springer Verlag. pp.
161-168.

\vspace*{4mm}

Jiang, Y.X., A.Lee, J.Y. Chen, M. Cadene, B. T. Chait, R. MacKinnon. 2002. The
open pore conformation of potassium channels. {\it Nature}. 417: 523-526.

\vspace*{4mm}

Liebovitch,~L.S., and A.T.~Todorov. 1996. Using fractals and nonlinear dynamics
to determine the physical properties of ion channel proteins. {\it Crit. Rev.
Neurobiology}. 10: 169-187.

\vspace*{4mm}

Mercik,~Sz., K.~Weron, and Z.~Siwy. 1999. Statistical analysis of ionic current
fluctuations in membrane channels. {\it Phys. Rev. E}. 60: 7343-7348.

\vspace*{4mm}

Mercik,~Sz., and K.~Weron. 2001. Stochastic origins of the long-range
correlations of ionic current fluctuations in membrane channels. {\it Phys. Rev.
E}. 63: 051910 (10 pages).

\vspace*{4mm}

Neher,~E. and B.~Sakmann. 1976. Single-channel current recorded from membrane of
denervated frog muscle fibers. {\it Nature}. 260: 799-802.

\vspace*{4mm}

Renner,~Ch., J.~Peinke, and R.~Friedrich. 2001. Markov properties of high
frequency exchange rate data. {\it Physica A}. 298: 499-520.

\vspace*{4mm}

Risken,~H. 1984. The Fokker-Planck Equation. Springer-Verlag, Berlin.

\vspace*{4mm}

Roncaglia,~R., R.~Mannella, and P.~Grigolini. 1994. Fractal Properties of Ionic
Channels and Diffusion. {\it Math. Biosci}. 123: 77-101).

\vspace*{4mm}

Schrempf,~H., O.~Schmidt, R.~Kuemmerlen, S.~Hinnah, D.~Mueller, M.~Betzler,
T.~Steinkamp, and R.~Wagner. 1995. TI A prokaryotic potassium ion channel with
two predicted transmembrane segments from Streptomyces lividans, {\it EMBO J.}
14: 5170-5178.

\vspace*{4mm}

Siwy,~Z., Sz.~Mercik, K.~Weron and M.~Ausloos. 2001. Application of dwell-time
series in studies of long-range correlation in single channel ion transport:
analysis of ion current through a big conductance locust potassium channel. {\it
Physica A}. 297: 79-96.

\vspace*{4mm}

Siwy,~Z., M.~Ausloos, and K.~Ivanova. 2002. Correlation studies of open and
closed states fluctuations in an ion channel. Analysis of ion current through a
large conductance locust potassium channel, {\it Phys. Rev. E}. 65: 031907 (6
pages).

\vspace*{4mm}

Siwy, Z., and A. Fulinski. 2002.  Origins of 1/f$^{\alpha}$ noise in membrane
channels currents. {\it Phys. Rev. Lett}. 89: 158101 (4 pages).

\vspace*{4mm}

Wagner, M.  and J. Timmer. 2000. The Effects of Non-Identifiability on Testing
for Detailed Balance in Aggregated Markov Models for Ion-Channel Gating. {\it
Biophys. J.}, 79: 2918-2924.


\newpage \begin{center}

{\Large Figure Captions}

\end{center}

\vspace{5mm}

{\bf Figure 1.} --- Locust potassium channel probability densities (PDF)
$p(\Delta T,\Delta n)$ of the dwell-times increments $\Delta T$ of (a) all dwell
times, i.e. open and closed; (b) same as (a) but with each pdf displayed
vertically to enhance the tail behaviour, all data (30604 data points) for the
time lags $\Delta n$, as indicated in the figure. The discretization step of the
histogram is 6 ms. (c) the portion between $\Delta T$=-30 and +30 ms is zoom out
with discretization step of 0.6 ms and each pdf displayed vertically to enhance
the tail behaviour, all data (30604 data points)  for the time lags $\Delta n$,
as indicated in the figure

\vspace*{5mm}

{\bf Figure 2.} --- Locust potassium channel probability densities (PDF)
$p(\Delta T,\Delta n)$ of the dwell-times increments $\Delta T$ of a closed
channel; (b) same as (a) but with each pdf displayed vertically to enhance the
tail behaviour, all data (15302 data points )  for the time lags $\Delta n$, as
indicated in the figure. The discretization step of the histogram is 6 ms. (c)
the portion between $\Delta T$=-30 and +30 ms is zoom out with discretization
step of 0.6~ms and each pdf displayed vertically to enhance the tail behaviour,
all data (15302 data points)  for the time lags $\Delta n$, as indicated in the
figure

\vspace*{5mm}

{\bf Figure 3.} --- Locust potassium channel probability densities (PDF)
$p(\Delta T,\Delta n)$ of the dwell-times increments $\Delta T$ of an open
channel; (b) same as (a) but with each pdf displayed vertically to enhance the
tail behaviour, all data (15302 data points )  for the time lags $\Delta n$, as
indicated in the figure. The discretization step of the histogram is 0.6~ms

\vspace*{5mm}

{\bf Figure 4.} --- Equal probability contour plots of the joint pdf for the
simultaneous occurrence of the dwell-times differences $p(\Delta T_{2},\Delta
n_{2},\Delta T_{1},\Delta n_{1})$ for two values of $\Delta n$, $\Delta n_1$ = 16
$\Delta n_0$,  $\Delta n_2$= 8 $\Delta n_0$: (a) for all dwell times; (b) for the
series consisting of durations of   closed, and (c) for the series consisting of
durations of  open states. Contour levels correspond to $log_{10}p$=-1, -1.5,
2,-2.5, -3,-3.5 from center to border

\vspace*{5mm}

{\bf Figure 5.} --- (a) Equal probability contour plots of the conditional pdf
$p(\Delta T_{2},\Delta n_{2} |  \Delta T_{1},$ $\Delta n_{1})$ for two values of
$\Delta n$, $\Delta n_1$ = 8 $\Delta n_0$,  $\Delta n_2$= $\Delta n_0$=0.1 ms,
for open-closed durations signal. Contour levels correspond to $log_{10}p$=-1,
2,-3, -4,  from center to border; (b) and (c) experimental and solution of the
Chapman Kolmogorov equation for the corresponding pdf at $\Delta T_1$ = -3 and
+3~ms

\vspace*{5mm}

{\bf Figure 6.} --- Kramers-Moyal drift and diffusion coefficients $D^{(1)}$ and
$D^{(2)}$ as a function of $\Delta T$ for the time series consisting of all
dwell-times. The quadratic fit of the diffusion coefficient is done for $\Delta T
< 14$~ms. $\Delta T$ changes with step 2~ms

\vspace*{5mm}

{\bf Figure 7.} --- Kramers-Moyal drift and diffusion coefficients $D^{(1)}$ and
$D^{(2)}$ as a function of $\Delta T$ for the time series consisting of closed
time intervals. The quadratic fit of the diffusion coefficient is done for
$\Delta T$ between -9~ms and +10~ms. $\Delta T$ changes with step 1~ms

\vspace*{5mm}

{\bf Figure 8.} --- Kramers-Moyal drift and diffusion coefficients $D^{(1)}$ and
$D^{(2)}$ as a function of $\Delta T$ for the time series consisting of  open
time intervals. The quadratic fit of the diffusion coefficient is done for
$\Delta T < 3$~ms and the linear fit is done for $\Delta T > -3$~ms. The linear
fits for $D^{(1)}$ are obtained for  $-3<\Delta T <0$~ms and  $0<\Delta T <3$~ms.
$\Delta T$ changes with step 0.2~ms
 
\end{document}